# Faint Blue Galaxies as a Probe of the X-ray Background at High Redshift

Marie A. Treyer & Ofer Lahav
*Institute of Astronomy, Madingley Road, Cambridge CB3 0HA, U.K.*

**ABSTRACT**
We present a formalism describing the physical content of cross-correlation functions between a diffuse background and a population of discrete sources. The formalism is used to interpret cross-correlation signals between the unresolved X-ray background and a galaxy population resolved to high redshift in another spectral band. Specifically, we apply it to the so-called faint blue galaxy population and constrain their X-ray emissivity and clustering properties. A model is presented which satisfies the recently measured constraints on all 3 correlation functions (galaxy/galaxy, background/background and galaxy/background). This model predicts that faint galaxies in the magnitude range $B = 18 - 23$ (covering redshifts $z \lesssim 0.5$) make up $\sim 22\%$ of the X-ray background in the $0.5 - 2$ keV band. At the mean redshift of the galaxy sample, $\bar{z} = 0.26$, the comoving volume emissivity is $\rho_X \sim 6 - 9 \times 10^{38} h$ ergs s$^{-1}$Mpc$^{-3}$. When extrapolated to fainter magnitudes, the faint blue galaxy population can account for most of the residual background at soft energy. We show how the measurement of the angular and zero-lag cross-correlation functions between increasingly faint galaxies and the X-ray background can in principle allow us to map the X-ray emissivity as a function of redshift.

## 1 INTRODUCTION

The X-ray background (XRB) has been the first diffuse cosmic radiation ever detected over three decades ago (Giacconi et al. 1962), but unlike the more famous microwave background discovered shortly after, its puzzling origin is yet to be solved. Although it is now known (e.g. Shanks et al. 1991) that quasars contribute a significant ($\gtrsim 30$ %) fraction of the soft X-ray part of the radiation ($\sim 1$ keV), the observed steep spectra of these objects, or of any others, do not resemble that of the hard X-ray emission.

In recent years the XRB has been studied by means of analyzing the total intensity, the spectrum and the spatial fluctuations (e.g. Boldt 1987 and Fabian & Barcons 1992 for reviews). In particular, the spatial fluctuations were analyzed by: (i) Source identification of high-flux regions (e.g. Shanks et al. 1991) (ii) Auto-correlation function (e.g. de Zotti et al. 1990, Carrera et al. 1993) (iii) Cross-correlation of the XRB with nearby galaxies (Turner & Geller 1980, Jahoda et al. 1991, Lahav et al. 1993, Miyaji et al. 1994, Carrera et al. 1995, Barcons et al. 1995).

In this paper we extend the cross-correlation approach to higher redshift and confront it with recent observations (Roche et al. 1995). We present a general formalism which takes into account the clustering of the sources, their evolution, and the world geometry. We then apply this formalism to the so-called faint blue galaxy population. The unexpectedly numerous sources observed at faint blue magnitude (Tyson 1988, Lilly et al. 1991, Metcalfe et al. 1992, Lilly 1993) have been a subject of active research in the last few years. It now seems that most of the 'excess' galaxies (with respect to the volume density expected from local surveys), are narrow emission-line objects at moderate redshift ($z < 1$), with star-forming or Seyfert activity (Broadhurst et al. 1992, Tresse et al. 1994). The occurrence of close pairs and disturbed morphologies among them (Glazebrook et al. 1995b, Driver et al. 1995) suggests that these object could be in the process of merging, thus explaining their much lower number density in local surveys. Or else, they may have declined in luminosity rather than number due the short time scale of their activity (e.g. Babul & Rees 1992). Whatever the actual process of their apparent vanishing, their starburst and/or Seyfert activity makes them likely X-ray sources, and therefore an XRB candidate population given their extremely large surface density on the sky (about $10^4$ objects per square degree at $B \sim 23$).

In the next section we review the necessary ingredients of cross-correlation analysis, namely the basic statistical properties of the galaxies, of the XRB and of the source clustering. In the third section, we write the cross-correlation and auto-correlation functions as a function of these ingredients, both in Fourier space in terms of power-spectrum, and in real space using more specific assumptions and approximations. These functions have been explicitly derived in Appendix A. We then constrain the parameters (Section 4) using a recent measurement of the cross-correlation function between a magnitude limited sample of faint blue galaxies and the soft XRB (Roche et al. 1995), and present a model for the faint blue galaxy contribution to the XRB. We show that X-ray emitters (such as starburst or Seyfert galaxies) among the faint galaxies can account for the observed statistical properties of the XRB and propose further statistics to quantify their X-ray contribution as a function of redshift. Our conclusions and future work are discussed in section 5.

## 2 BASIC PROPERTIES AND ASSUMPTIONS

In this study, we cross-correlate known galaxies with the diffuse XRB. The calculations involve properties of the galaxies, of the unknown X-ray sources, and their spatial clustering.

### 2.1 The galaxies

Let us assume a population characterized by a luminosity function $\phi_\nu(L_\nu, z)$ at frequency $\nu$ and redshift $z$, and a spectral energy distribution of the form $L_\nu \propto \nu^{-\beta}$ (where $\beta$ is the spectral index in the relevant band $\nu \pm \Delta\nu$). In a survey with flux limits $f_{min}$ and $f_{max}$ at frequency $\nu$, the number density of resolved galaxies at redshift $z$ will be:

$$n(z) = \int_{L_{min}(z)}^{L_{max}(z)} \phi_\nu(L_\nu, z) dL_\nu, \qquad (1)$$

where $L_{min,max}(z) = 4\pi r_l^2(z)(1+z)^{\beta-1} f_{min,max}$ and $r_l(z)$ is the luminosity distance. We call $\rho_\nu(z)$ the total volume emissivity of the population *observed* in the band $\nu \pm \Delta\nu$, *i.e.* originating from the frequency band $(1+z)\nu \pm (1+z)\Delta\nu$:

$$\rho_\nu(z) = (1+z)^{-\beta+1} \int_0^\infty \phi_\nu(L_\nu, z) L_\nu dL_\nu. \qquad (2)$$

In the case of optically selected galaxies, the observables are: (i) the number of galaxies $N(m,z)dmdz$ per deg$^2$ on the sky at magnitude $m$ and redshift $z$, measured from deep spectroscopic surveys down to an optical magnitude of $B \approx 24$ (e.g. Glazebrook et al. 1994a), and (ii) the galaxy number counts $C(m) = \int_z N(m,z) dz$, observed to $B \approx 27$ (e.g. Tyson 1988). We use the following fit to the galaxy redshift distribution in the $B_J$-band proposed by Efstathiou (1995):

$$N(m,z) = C(m) \frac{3z^2}{2 z_c^3(m)} \exp\left[-\left(\frac{z}{z_c(m)}\right)^{3/2}\right], \qquad (3)$$

where:

$$z_c(m) = \begin{cases} 0.0113(m-17)^{1.5} + 0.0325 & \text{if } 17 \lesssim m \lesssim 22, \\ 0.0010(m-17)^3 + 0.0325 & \text{if } m \gtrsim 22. \end{cases}$$

The evolution of the galaxy luminosity function results in a steepening of the faint end slope, *i.e.* an increase in the number or/and luminosity of locally faint and blue galaxies. The volume density $n(z)$ of catalogued galaxies at redshift $z$ (Eq. 1) appearing in the correlation functions verifies:

**Table 1.** Parameters on the X-ray background and X-ray sources:

|  | Soft band 0.5-2 keV (ROSAT) | Hard band 2-10 keV (HEAO1) |
|---|---|---|
| $\bar{I}$ | $\sim 1.3 - 2.1 \times 10^{-8}$ | $5.2 \times 10^{-8}$ ergs s$^{-1}$cm$^{-2}$sr$^{-1}$ |
| $\beta_{XRB}$ | $0.4 - 1.1$ | $0.4$ |
| $\rho_0$ | ? | $\sim 10^{39} h$ ergs s$^{-1}$Mpc$^{-3}$ |
| $q_{AGN}$ | $\sim 3$ | ? |
| $\beta_{AGN}$ | $\sim 0.7 - 1$ | $\sim 0.7 - 1$ |

$$n(z) dv(z) = dz \int_{B_{min}}^{B_{max}} N(m,z) dm, \qquad (4)$$

where $dv(z)$ is the volume element per unit solid angle at redshift $z$ (Appendix B). The mean$^\star$ projected number of galaxies per unit solid angle is:

$$\bar{N} = \int_z n(z) dv(z) = \int_{B_{min}}^{B_{max}} C(m) dm. \qquad (5)$$

### 2.2 The X-ray background

Resolved X-ray sources, essentially AGN's with $\sim 1$ keV-fluxes $\gtrsim 10^{-14}$ ergs s$^{-1}$cm$^{-2}$, contribute $\sim 30$ % of the total extragalactic X-ray intensity (Hasinger 1992, Shanks et al. 1991). In this paper, the 'X-ray background' refers to the *unresolved* fraction of this radiation. Let us call $\rho_x(z) = \rho_d(z) + \rho_s(z)$ the volume emissivity of the XRB due to diffuse radiation and unresolved discrete sources at redshift $z$, observed in a given X-ray band. The diffuse component $\rho_d$, first thought to be dominant and due to hot inter-galactic gas, has since been shown to be strongly limited. Too much hot gas in the Universe would indeed distort the microwave background spectrum, and the COBE experiment has proved that no such effect was observed to a very high accuracy (Mather et al. 1990). The discrete source component $\rho_s$ is therefore strongly dominant, although no known X-ray source population has yet been shown fully consistent with the background properties. $\rho_s$ may be writen as the added contributions of various populations emitting in this band, *i.e.* a sum of Eq. 2's. In the following, we will assume the diffuse component of the XRB to be negligible ($\rho_x(z) \approx \rho_s(z)$), and model the *observed* X-ray light density of the sources in a given X-ray band as:

$$\rho_s(z) = \rho_0 (1+z)^q. \qquad (6)$$

The mean intensity of the XRB per unit solid angle is:

$$\bar{I} = \int_z \frac{\rho_x(z)}{4\pi r_l^2(z)} dv(z) = \frac{\rho_0}{4\pi} \int_z (1+z)^{q-2} F dr_c, \qquad (7)$$

where $F dr_c = dv/r_c^2$ is a geometry dependent function of $z$ defined in Appendix B. The soft (E $\sim 1-2$ keV) and hard (E $\gtrsim 2$ keV) XRB exhibit vary distinct properties (see Boldt 1987 for a review) and we actually possess very different data depending on the energy band. The known (but somewhat uncertain) parameters of the XRB and sources (based on Boldt 1987, Fabian & Barcons 1992, Hasinger et al. 1993, Boyle 1994, Chen et al. 1994) are summarized in Table 1. The measured intensity in the soft band actually varies within a factor of almost 2, depending (among other things) on the Galactic component and resolved source removal procedures. In the following sections, we shall use $\bar{I} = i \times 10^{-8}$ ergs s$^{-1}$cm$^{-2}$sr$^{-1}$, where $1.3 \lesssim i \lesssim 2.1$ in the $0.5 - 2$ keV energy band.

---

$^\star$ Means values will be indicated as either $\bar{X}$ or $<X>$.

### 2.3 Galaxies and X-ray sources

We shall characterize the cross-correlation function between the hypothetical X-ray sources and the known galaxies by the usual function $\xi(r, z)$. The excess probability of simultaneously finding a galaxy and an X-ray source in the volume elements $\delta V_1$ and $\delta V_2$ respectively, with separation $r_{12}$, is defined as:

$$< n(\delta V_1) \rho_s(\delta V_2) > = [1 + \xi(r_{12}, z)] n(z) \delta V_1 \rho_s(z) \delta V_2. \quad (8)$$

The spatial auto-correlation functions of nearby populations (e.g. galaxies and clusters) are well approximated by power-laws. For the cross-correlation function between the X-ray sources and the faint galaxies we shall assume a similar form, taking into account clustering evolution with redshift (Peebles 1980):

$$\xi(r, z) = (1+z)^{-(3+\epsilon)} \left(\frac{r}{r_o}\right)^{-\gamma}, \quad (9)$$

where $r$ is the *physical* separation between the sources and $\epsilon$ an arbitrary evolution parameter. Theoretical models commonly assume $\gamma - 3 \leq \epsilon \leq 0$ for the galaxies. The upper limit corresponds to 'stable clustering' while in the lower limit case, the growth of density perturbations is due to the expansion of the Universe only. For local bright galaxies, the correlation function parameters are $\gamma \sim 1.8$ and $r_0 \sim 5\ h^{-1}$Mpc (Peebles 1980). Faint galaxies on the other hand seem to be significantly less strongly clustered, with $r_0 \sim 2\ h^{-1}$Mpc (Efstathiou 1995).

To allow for non power-law models and more direct comparison with theoretical models, it is convenient to express the cross-correlation in terms of power spectrum in Fourier space:

$$P(k, z) = \frac{P(k)}{(1+z)^{\epsilon'}}, \quad (10)$$

where $\epsilon'$ is an evolution parameter. (In the case where the power spectrum is a power-law with slope $n$, $\epsilon' = \epsilon - n$). As a working hypothesis, we assume all three correlation functions to be the same (as would be the case for instance if the faint galaxies were the only clustered X-ray sources), but the formalism can easily be modified to allow the auto and cross-correlation functions to be different.

Operationally, the statistical angular galaxy/background cross-correlation is performed by dividing the sky into cells of small solid angle and calculating :

$$\eta_{gb}(\theta) = < (N - \bar{N})_{\theta'} (I - \bar{I})_{\theta'+\theta} >_{cells}, \quad (11)$$

or, in a normalized form:

$$w_{gb}(\theta) = \frac{\eta_{gb}(\theta)}{\bar{N}\bar{I}}, \quad (12)$$

where the average is over all pairs of cells with angular separation $\theta$.

We use the standard Robertson-Walker metric relations for the world geometry (distances, volume, etc.). These relations are recalled in Appendix B. Unless otherwise stated, we assume an $\Omega = 1, \Lambda = 0$ universe, and write the Hubble constant as $H_0 = 100\ h$ km/s/Mpc.

## 3 CORRELATION FUNCTIONS

We have derived the formalism required to interpret cross-correlation and auto-correlation analysis. (Details of these somewhat cumbersome calculations are given in Appendix A to avoid overloading the main text). We first write the general formulae in Fourier space and then propose a more practical version in real space, applied to the faint blue galaxy population.

### 3.1 Angular correlation functions

Given two functions of galaxy positions, $X$ and $Y$, the angular cross-correlation function between them at angular separation $\theta$ can be written as:

$$\eta_{XY}(\theta) = \int_0^\infty P(k, z)\ k\ \tilde{g}_{XY}(k\theta)\ \mathrm{d}k, \quad (13)$$

where $P(k, z)$ is the power spectrum at redshift $z$ and $\tilde{g}_{XY}$ is an appropriate window function of $X$ and $Y$. For the galaxy/galaxy and background/background angular auto-correlation functions respectively:

$$\tilde{g}_{gg}(k\theta) = \frac{1}{2\pi} \int_z n^2(z) J_0(kr_c\theta) r_c^4 F \mathrm{d}r_c \quad (14)$$

$$\tilde{g}_{bb}(k\theta) = \frac{1}{2\pi} \int \left(\frac{\rho_s(z)}{4\pi r_l^2}\right)^2 J_0(kr_c\theta) r_c^4 F \mathrm{d}r_c. \quad (15)$$

For the galaxy/background angular cross-correlation function:

$$\tilde{g}_{gb}(k\theta) = \frac{1}{2\pi} \int_z n(z) \frac{\rho_s(z)}{4\pi r_l^2} J_0(kr_c\theta) r_c^4 F \mathrm{d}r_c. \quad (16)$$

In the above and following equations, $r_l$, $r_c$, $\mathrm{d}v$ and $F$ are functions of $z$ defined in Appendix B. We omit to write the $z$ dependency in order to make the equations more compact. $J_0$ is a Bessel function of order zero, the definition of which appears in Appendix A. To illustrate the scales probed by the three statistical measures as a function of angular scale, we have plotted the three window functions $\tilde{g}_{XY}$ (assuming an arbitrary set of parameters described in the caption) for $\theta = 1'$ (upper three lines) and $\theta = 1°$ (lower three lines). The thick solid and dashed lines represent $P(k)k^2$ for a standard CDM model and a low density CDM model respectively. The product $P(k)k^2 \times \tilde{g}_{XY}$ is the contribution per logarithmic interval $\mathrm{d}\ln k$ centered on wave number $k$ to the integral in Eq. 13. On the arcminute scale, the kernels are constant with $k$ for all three statistics down to very small scales ($k^{-1} \sim 0.1\ h^{-1}$Mpc). As they are functions of $k\theta$, increasing $\theta$ simply results in shifting the kernels towards smaller $k$. For example on the degree scale, the three statistics are less sensitive to scales smaller than $k^{-1} \sim 10\ h^{-1}$Mpc. Note that the XRB statistics is flux-weighted ($\propto r_l^{-2}$), hence less sensitive to larger scales compared with the number-weighted galaxy statistics. However, as in our model the XRB sources are distributed over a much larger range of redshifts, the background auto-correlation probes larger wavelenghts than the galaxy auto-correlation (and the galaxy/background cross-correlation probes intermediate scales).

FIGURE 1

### 3.2 Zero-lag cross-correlation function

At zero-lag ($\theta = 0$), we consider cells of solid angle $\omega$. A Poisson term, due to the potential X-ray emissivity $\rho_g$ of the catalogued galaxies themselves, adds to the clustering term:

$$\eta_0 = \eta_p + \eta_c. \tag{17}$$

The poisson term is defined as:

$$\eta_p = \omega \Delta \bar{I}_g = \omega \int_z \frac{\rho_g(z)}{4\pi r_l^2} dv. \tag{18}$$

where $\Delta \bar{I}_g$ is the X-ray intensity contributed by the catalogued galaxies. The clustering term can be writen:

$$\eta_c = \int_0^\infty P(k,z) \, k \, \tilde{g}_0(k) \, dk, \tag{19}$$

where the kernel $\tilde{g}_0$ is given by:

$$\tilde{g}_0(k) = \frac{1}{2\pi} \int n(z) \frac{\rho_s(z)}{4\pi r_l^2} \int_\omega \int_\omega J_0(kr_c\theta) d\Omega_1 d\Omega_2 \, r_c^4 F \, dr_c.$$

### 3.3 Modeling

With the various prescriptions and approximations described in Section 2 and Appendix A, the angular correlation functions result in simple power-laws of the angular separation $\theta$ with power index $1 - \gamma$. We can write all three functions in the form:

$$\eta_{XY}(\theta) = A_{XY} \theta^{1-\gamma}, \tag{20}$$

where:

$$A_{gg} = H_\gamma r_0^\gamma \int_z dz (1+z)^p \frac{r_c^{1-\gamma}}{F \frac{dr_c}{dz}} \left( \int_{B_{min}}^{B_{max}} N(m,z) dm \right)^2 \tag{21}$$

$$A_{bb} = H_\gamma r_0^\gamma \left( \frac{\rho_0}{4\pi} \right)^2 \int_z dz (1+z)^{p+2(q-2)} r_c^{1-\gamma} F \frac{dr_c}{dz} \tag{22}$$

$$A_{gb} = H_\gamma r_0^\gamma \frac{\rho_0}{4\pi} \times$$

$$\int_z dz (1+z)^{p+q-2} r_c^{1-\gamma} \int_{B_{min}}^{B_{max}} N(m,z) dm \tag{23}$$

where $p = \gamma - (3+\epsilon)$ is a parameter describing the clustering properties of the galaxies with the XRB sources. Assuming for simplicity square cells with size $\omega = a^2$, the clustering term (Eq. 19) in the expression of the zero-lag cross-correlation can be written (Appendix A):

$$\eta_c = C_\gamma A_{gb} a^{5-\gamma}. \tag{24}$$

For $\gamma = 1.8$, $H_\gamma = 3.68$ and $C_\gamma = 2.25$. The parameters we are interested in essentially are: the volume emissivity of the X-ray sources as a function of redshift, parameterized by $\rho_0$ and $q$, and the clustering parameters $r_0$ and $p$. The autocorrelation functions of both the galaxies and the XRB have been studied elsewhere by various authors (e.g. Peebles 1980, de Zotti et al. 1990, Carrera et al. 1993). Our purpose here is to emphasize the cross-correlation techniques between high redshift galaxies and the diffuse background. The faint blue galaxies selected in a given magnitude range, stand as beacons allowing us to probe the XRB in the particular redshift range these galaxies are predominantly populating. Thus, measuring the amplitude $A_{gb}$ of the cross-correlation function in various magnitude ranges can provide constraints on the clustering parameters and the volume emissivity of the X-ray sources.

## 4 THE FAINT GALAXY CONTRIBUTION TO THE XRB

### 4.1 Constraining the parameter space

We have fitted the measured angular cross-correlation between the XRB in the $0.5 - 2$ keV band and faint galaxies in the magnitude range $B = 18 - 23$ (Roche et al. 1995) by the following power index and amplitude:

$$\begin{aligned} \gamma &= 1.85 \pm 0.05 \\ A_* &= (1.5 \pm 0.4) \times 10^{-5} \times \bar{I}\bar{N} \end{aligned} \tag{25}$$

The observed values for the mean intensity of the unresolved XRB in the above energy band and for the mean number of $18 < B < 23$ galaxies are taken to be: $\bar{I} = i \times 10^{-8}$ ergs s$^{-1}$cm$^{-2}$sr$^{-1}$ where $1.3 \lesssim i \lesssim 2.1$, and $\bar{N} = 2.6 \times 10^7$ galaxies per steradian respectively. For the purpose of exploring the parameter space, we introduce $\aleph = p + q - 2 = \gamma - \epsilon + q - 5$ as a global evolution parameter, and define the function:

$$f(\aleph) = \int_z (1+z)^\aleph r_c^{1-\gamma} \int_{B=18}^{B=23} N(m,z) dm. \tag{26}$$

(We assume $B = B_J + 0.2$). With these new parameters, we can rewrite Eq. 23 in the following compact parametric form:

$$\rho_0 r_0^\gamma = \frac{4\pi A_*}{H_\gamma f(\aleph)}, \tag{27}$$

which we use to constrain the background emissivity and the clustering length. We consider a range of theoretical evolution models for the clustering of the galaxies with the X-ray sources: $-1.2 < \epsilon < 0$, and a range of evolution for the light density of the X-ray sources: $0 < q < 4$ (from no-evolution to quasar-like evolution). These ranges imply $-3 < \aleph < 3$, which we assume to be a reasonable and wide enough range for this parameter. Figure 2a shows $r_0^\gamma \rho_0$ as a function of $\aleph$ (Eq. 27).

FIGURE 2a

We find: $6 \times 10^{38} \lesssim \rho_0 r_0^\gamma \lesssim 17 \times 10^{38}$ (where $r_0$ is in $h^{-1}$Mpc and $\rho_0$ is in $ih$ ergs s$^{-1}$Mpc$^{-3}$). Assuming $2 \leq r_0 \leq 6$ $h^{-1}$Mpc and $1.8 \leq \gamma \leq 1.9$, this result implies:

$$2.3 \times 10^{37} \lesssim \rho_0 \lesssim 4.9 \times 10^{38} \, ih \text{ ergs s}^{-1}\text{Mpc}^{-3}. \tag{28}$$

It is clear that measuring the angular cross-correlation function alone is insufficient to deduce the X-ray emission of the galaxies, unless very specific assumptions are made. Assuming as we did that the faint galaxies essentially cluster with themselves, *i.e.* that the cross-correlation between the background and the galaxies arises from their auto-correlation as opposed to their potential clustering with other X-ray sources, $\rho_0$ parameterizes the galaxy own X-ray emissivity. In this case the background auto-correlation function (ACF) derived in Eq. 22 is the galaxy contribution to the XRB fluctuations. (Other clustered populations may produce additional fluctuations). Recent determinations (Chen et al. 1994) [†] of the soft XRB anisotropy set the following 2-$\sigma$ upper limit:

$$w_{bb}(\theta) \leq 3.86 \times 10^{-3} \left(\frac{\theta}{1'}\right)^{0.8}, \qquad (29)$$

on scales $\theta \sim 2 - 300$ arcminutes, or, in the present notation (Eqs. 20 and 22):

$$A_{bb} \leq 9.6 \times 10^{-6} \bar{I}^2. \qquad (30)$$

Using both auto and cross-correlation measurements and combining Eqs. 22 and 27, we derive another constraint on $\rho_0$ (independent of $r_0$) as a function of the evolution parameters:

$$\rho_0 \leq 1.2 \times 10^{-4} \frac{A_*^{-1} \bar{I}^2 f(p+q-2)}{\int_z (1+z)^{p+2(q-2)} r_c^{1-\gamma} F \mathrm{d}r_c}, \qquad (31)$$

where $f$ is the function defined in Eq. 26. The two solid lines in Fig. 2b bracket the above upper limit as a function of $q$ when the parameter $p = \gamma - \epsilon - 3$ varies from $-1.2$ (upper line) to 0 (lower line). (The error bars on $A_*$ are ignored in this calculation).

FIGURE 2b

### 4.2 The faint blue galaxy contribution

The galaxy auto-correlation function $\eta_{gg}$ has been measured by several groups. At faint magnitude, the data are consistent with low $r_0$ in contrast with the observation of bright local galaxies (Efstathiou 1995). We assume the fitting values of $r_0 = 2\ h^{-1}$Mpc, $\gamma = 1.8$ and $\epsilon = -1.2$ proposed by Efstathiou (1995). For the spectral indices, we assume $\beta_B \sim 1$ — as the faint blue galaxies are predominantly late-type/Irr galaxies at increasing magnitude — and $\beta_X \sim 0.4$ appropriate for the spectrum of the unresolved XRB in both the soft and hard bands. Such a spectrum has been observed in absorbed AGN's and starburst galaxies and may therefore be appropriate for a large fraction of the faint blue galaxies. This parameter is actually barely significant when the others have been fixed. (Assuming $\beta_X = 1$ (bright AGN-like) changes the results below by less than 10%). The optical flux of a galaxy with magnitude $m$ is $F_B = 10^{-0.4(m-\kappa)}$, where $\kappa = 16.6$ for fluxes in units of $10^{38} h$ ergs s$^{-1}$Mpc$^{-3}$. The total optical emissivity due to galaxies (of all magnitudes) at redshift $z$ verifies:

---

[†] These measurements have been carried out in the 0.4-2.4 keV band whereas $\bar{I}$ in defined in the 0.5-2 keV band. Using the spectral parameters quoted by the authors, we apply a mean spectral correction of $\bar{I}_{0.4-2.4} = 1.3\ \bar{I}$.

$$\frac{\rho_B(z)}{4\pi r_l^2(z)} \mathrm{d}v(z) = \mathrm{d}z \int_{-\infty}^{+\infty} 10^{-0.4(m-\kappa)} N(m,z) \mathrm{d}m \qquad (32)$$

Our adopted redshit distribution, arbitrarily extrapolated to $B = 30$ to account for 'all' the galaxies, yields $\rho_B(z) \approx 3.8 \times 10^{41}(1+z)^{1.3}\ h$ ergs s$^{-1}$Mpc$^{-3}$ out to $z \approx 2$. The correlation between optical and X-ray luminosities is approximately linear for late-type galaxies, although very scattered (Fabbiano et al. 1988). Assuming $L_X \propto L_B$ at all redshifts and for all galaxies, then $\rho_X(z) \propto \rho_B(z)(1+z)^{\beta_B - \beta_X}$. In the present notation, this scenario corresponds to $q = 1.3 + \beta_B - \beta_X = 2$, and therefore $\aleph = \gamma - \epsilon + q - 5 = 0$. The above values of $\aleph$ and $r_0$ imply (Eq. 27):

$$\rho_0 \approx 2.7 \times 10^{38}\ ih\ \text{ergs s}^{-1}\text{Mpc}^{-3} \qquad (33)$$

for the volume emissivity of the galaxies at soft X-rays, and yield:

$$\log\left(\frac{L_X}{L_B}\right) = \log\left(\frac{\rho_0}{\rho_B(0)}\right) \approx -3.1 + \log(i), \qquad (34)$$

within the range of observed values for late-type galaxies (Fabbiano et al. 1988). This luminosity ratio implies that $B \gtrsim 18$ galaxies have soft X-ray fluxes fainter than $10^{-14}$ ergs s$^{-1}$cm$^2$. The source counts are shown in Fig. 3.

FIGURE 3

At the mean redshift of the sample $\bar{z} = 0.26$, the comoving volume emissivity contributed by those X-ray sources which are clustered with the galaxies is:

$$\rho_X(\bar{z}) = \rho_0(1+\bar{z})^q \sim 4.2 \times 10^{38}\ ih\ \text{ergs s}^{-1}\text{Mpc}^{-3}. \qquad (35)$$

The integrated X-ray flux from $18 < B < 23$ galaxies:

$$\Delta \bar{I}_g = \frac{L_X}{L_B} \times \qquad (36)$$

$$\int_z \mathrm{d}z (1+z)^{\beta_B - \beta_X} \int_{B=18}^{B=23} 10^{-0.4(m-\kappa)} N(m,z)\mathrm{d}m$$

represents $\sim 22\%$ of the XRB in the ROSAT band. The integrated X-ray flux of 'all' (*i.e.* $B < 30$) galaxies depends on the maximum redshift of integration, and obviously on our arbitrary extrapolation of the redshift distribution above $B \sim 24$. The 'total' faint blue galaxy population, as modeled here, may contribute anywhere from 55 to 75 % of the soft unresolved XRB as $z_{max}$ increases from 2 to 5. Figure 4 shows the predicted correlation functions. The model is consistent with the upper limit of the auto-correlation function of the background, although a slightly larger value for $r_0 = 3\ h^{-1}$Mpc would improve the fit, as well as decrease the contribution of the catalogued galaxies to $\sim 10\%$.

FIGURE 4

### 4.3 Further statistics

Measuring the amplitude of the cross-correlation function between the diffuse XRB and increasingly faint galaxies can, in theory, allow us to significantly narrow down the parameter space, and eventually derive fairly good estimates for the volume emissivity and the clustering properties of the X-ray sources at increasing redshift.

$$\delta A_{gb} = H_\gamma r_0^\gamma \frac{\rho_0}{4\pi} \int_z (1+z)^\aleph r_c^{1-\gamma} N(m,z) \mathrm{d}z \, \delta m$$

$$= \frac{A_*}{f(\aleph)} \int_z (1+z)^\aleph r_c^{1-\gamma} N(m,z) \mathrm{d}z \, \delta m \quad (37)$$

where $f(\aleph)$ is the function defined in Eq. 26. Figure 5a shows the normalized amplitude $\delta w_{gb}(m) = \delta A_{gb}(m)/\bar{I}C(m)$ as a function of $B$-magnitude when $\aleph$ varies from $-3$ to 3.

FIGURE 5a

Measuring the cross-correlation function at zero-lag can, again in theory, yield the galaxy contribution to the background as a function of magnitude ($\delta \bar{I}_g(m)$) in a straightforward way. Equations 17, 18 and 24 imply:

$$\delta \eta_0(a) = a^2 \delta \bar{I}_g(m) + a^{5-\gamma} C_\gamma \delta A_{gb}(m) \quad (38)$$

where the cross-correlated cells are assumed to be square solid angles $a^2$. Consequently we can hope to derive the X-ray contribution of a flux limited sample as well as the amplitude of the angular correlation function, by measuring $\eta_0$ for different cell sizes. In the magnitude range of the data:

$$\eta_0(a) = a^2 \Delta \bar{I}_g + a^{5-\gamma} C_\gamma A_*. \quad (39)$$

Fig.5b shows the lower and upper limits of the normalized zero-lag cross-correlation functions: $w_0 = \eta_0/a^4 \bar{I} \bar{N}$, when $\Delta \bar{I}_g/\bar{I}$ varies from 10 to 100% of the soft XRB. The solid line is our model prediction, *i.e.* $\Delta \bar{I}_g/\bar{I} \sim 22\%$.

FIGURE 5b

## 5 CONCLUSIONS

This paper presents a formalism for the cross-correlation function between a diffuse background radiation, such as the XRB, and a population of discrete sources, such as the so-called faint blue galaxy population. By extending earlier modeling of the cross correlation of the XRB with galaxies at zero redshift, the present formalism allows to constrain the X-ray emissivity and clustering properties of these faint galaxies and other X-ray sources clustered with them.

Cross correlation analyses of the hard ($\sim 2-10$ keV) XRB with local ($z \lesssim 0.03$) optical and IRAS galaxies yielded $\rho_0 \sim 0.7-1.2 \times 10^{39} h$ ergs/Mpc$^3$ (Lahav et al. 1993; Miyaji et al. 1994; Carrera et al. 1995). However, these results did not constrain the contribution to the XRB from high redshift. Roche at al. (1995) have recently measured the cross-correlation function between faint galaxies in the magnitude range $B = 18-23$ and the soft ($\sim 1$ keV) X-ray background, but only provided a heuristic interpretation to their measurement. An application of our formalism to a specific model predicts that these $18 < B < 23$ faint galaxies (covering redshifts $z \lesssim 0.5$) make up $\sim 22\%$ of the 1 keV XRB. When extrapolated to fainter magnitudes, the faint blue galaxy population can account for most of the residual background at soft energy. Our model satisfies all the available constraints on the 3 correlation functions (galaxy/galaxy, background/background and galaxy/background). However, we do not consider here the so-called 'spectral paradox', *i.e.* that the spectrum of the XRB (in both soft and hard bands) is not compatible with the spectrum of known sources. It remains an observational challenge (and a test for our model) to accurately measure the X-ray spectrum of the faint blue galaxies and other high redshift objects.

While the idea that the faint blue galaxies could significantly contribute to the XRB seems plausible, we wish to emphasize several uncertainties in the current measurements and modeling, and to propose other strategies. The contribution $\Delta \bar{I}_g/\bar{I} \sim 22\%$ we have derived for the $18 < B < 23$ galaxy population results from very specific assumptions. However appropriate these may be, different scenarios may lead to quite different conclusions. For example, a strong clustering model with $r_0 = 5 \, h^{-1}$Mpc (as observed for bright optical galaxies) and other previous assumptions unchanged, predicts $\Delta \bar{I}_g/\bar{I} \sim 4\%$. As the contribution drops, the model remains consistent with the background auto-correlation function. Inversely, very weak clustering with $r_0 = 1 \, h^{-1}$Mpc and $\epsilon = 0$, results in $\Delta \bar{I}_g/\bar{I} \sim 95\%$ of the background. Another option may be that the faint blue galaxies emit very negligible X-ray ($\Delta \bar{I}_g \approx 0$) but are strongly clustered with other sources of the XRB (e.g. very faint AGN's, or hot gas confined in poor galaxy groups), thus inducing a cross-correlation signal.

On the observational side, it seems that important quantities such as the mean soft XRB intensity and spectral index, are still uncertain, in part due to the different ways 'resolved' sources and the galactic component (strong below $\sim 1$ keV) are removed in various analyses. In deriving the non zero-lag cross-correlation it is also crucial to carefully take into account the detector's point-spread-function, in order to make sure that self-correlation of objects does not contribute to the signal. Regarding data analysis, the zero-lag cross-correlation is in fact easier to measure, taking into account the beam smearing (cf. Miyaji et al. 1994), and more informative than the non zero-lag correlation. It would also be interesting to apply these cross-correlation techniques to higher redshift objects (e.g. quasars and clusters) and in harder X-ray energy bands where most of the XRB energy resides, and to relate it to the emissivity derived from the cross-correlation with local galaxies. As demonstrated in this paper, the cross-correlation technique is a powerful probe of the X-ray emissivity as a function of redshift.

## ACKNOWLEDGEMENTS

We thank B. Boyle, L.-W. Chen, A. Dekel, A. Fabian, D. Helfand and T. Piran for stimulating discussions. We warmly acknowledge the hospitality of the Hebrew University in Jerusalem where a significant part of this work has been carried out. MAT is grateful to the European Community for a postdoctoral fellowship.

## REFERENCES

Babul A., Rees M., 1992, MNRAS, 255, 346
Barcons X., Franceschini A., de Zotti G, Danese L., Miyaji T., 1995, preprint.

Baugh C.M., Efstathiou G., 1993, MNRAS, 265, 145
Boldt E., 1987, Phys. Reports, 146, 215
Boyle B., 1994, in: The Nature of Compact Objects in Active Galactic Nuclei, Cambridge University Press
Broadhurst T., Ellis R., Glazebrook K., 1992, Nature, 355, 55
Carrera F.J., Barcons X., Butcher J.A., Fabian A.C., Stewart G.C., Toffolatti L., Warwick R.S., Hayashida K., Inoue H., Kondo H., 1993, MNRAS, 260, 376
Carrera F.J., Barcons X., Butcher J.A., Fabian A.C., Lahav O., Stewart G.C., Warwick R.S., 1995, ApJ, 275, 22
Chen L.-W., Fabian A.C., Warwick R.S., Branduardi-Raymont G., Barber C.R., 1994, MNRAS, 266, 846
Driver S., Windhorst R., Ostrander E., Keel W., Griffiths R., Ratnatunga K., 1995, ApJ, 449, L23
Efstathiou G., 1995, MNRAS, 272, L25
Fabbiano G., Gioia I., Trinchieri G., 1988, ApJ, 324, 749
Fabian A.C., Barcons X., 1992, ARA&A, 30, 429
Giacconi R., Gursky J., Paolini F., Rossi B., 1962, Phys. Rev. Let., 9, 439
Glazebrook K., Ellis R., Colless M., Broadhurst T., Allington-Smith J., Tanvir N., Taylor K., 1995a, MNRAS, 273, 157
Glazebrook K., Ellis R., Santiago B., Griffiths R., 1995b, MNRAS 275, L19
Hasinger G., 1992, In: The X–Ray Background, ed. Barcons X. & Fabian A.C., Cambridge University Press
Hasinger G., Burg R., Giacconi R., Hartner G., Schmidt M., Trümper J., Zamorani G., 1993, A&A, 271, 1
Jahoda K., Lahav O., Mushotzky R.F., Boldt E.A., 1991, ApJ, 378, L37; Erratum in 1992, ApJ, 399, L107
Lahav O., Fabian A.C., Barcons X., Boldt E., Butcher J., Carrera F.J., Jahoda K., Miyaji T., Stewart G.C., Warwick R.S., 1993, Nature, 364, 693
Lilly S., Cowie L., Gardner J., 1991, 369, 79
Lilly S., 1993, ApJ, 411, 501
Metcalfe N. et al., 1992, Texas/PASCOS Ann. NY Acad., 688, 934
Miyaji T., Lahav O., Jahoda K., Boldt E., 1994, ApJ, 434, 424
Peebles P.J.E., 1980, *The Large Scale Structure of the Universe*, Princeton University Press, Princeton
Roche N., Shanks T., Georgantoulos I., Stewart G.C., Boyle B.J., Griffiths R., 1995, MNRAS, 273, L15
Shanks T., Georgantopoulos I., Stewart G.C., Pounds K.A., Boyle B.J., Griffiths R.E., 1991, Nature, 353, 315
Totsuji H., Kihara T., 1969, PASJ, 21, 221
Tresse L., Rola C., Hammer F., Stasińska G., 1994, in: Wide Field Spectroscopy And The Distant Universe, eds. Maddox S. & Aragón-Salamanca A., World Scientific
Turner E.L., Geller M.J., 1980, ApJ, 236, 1
Tyson J.A., 1988, AJ, 96, 1
Wu X., Hamilton T., Helfand D.J., Wang Q., 1990, ApJ, 379, 564
de Zotti G. et al., 1990, ApJ, 351, 22

# APPENDIX A: ZERO-LAG CROSS-CORRELATION FUNCTION

We start by deriving the zero-lag cross-correlation between the galaxies and the background. The angular cross-correlation and auto-correlation functions follow quite simply from the same principles. We divide the sky into cells of small solid angle $\omega$ and calculate the following quantity:

$$\begin{aligned}\eta_0 &= <\delta N \delta I>_{cells} \\ &= <(N-\bar{N})(I-\bar{I})> \\ &= <NI> - <N><I>\end{aligned} \quad (A1)$$

where the average is over all cells. We assume the cells can be divided into elements $\delta V_k$ containing $n_k = 0$ or 1 galaxy (cf. Peebles 1980):

$$N = \sum n_k$$

$$I = \sum i_k, \quad \text{where} \quad i_k = \left(\rho_k/4\pi r_k^2\right)\delta V_k$$

$$NI = \sum n_k i_k + \sum_{j\neq k} n_j i_k$$

$$<NI> = \int \frac{\rho_g}{4\pi r_l^2}dV + \int\int n_1 \frac{\rho_2}{4\pi r_{l_2}^2}[1+\xi(r_{12})]dV_1 dV_2$$

$$<N><I> = \int n_1 dV_1 \int \frac{\rho_2}{4\pi r_{l_2}^2}dV_2$$

Therefore:

$$\eta_0 = \eta_p + \eta_c \quad (A2)$$

where $\eta_p$ is the poisson term and $\eta_c$ the clustering term defined as follows:

$$\eta_p = \int_{V_{cell}} \frac{\rho_g}{4\pi r_l^2}dV \quad (A3)$$

$$\eta_c = \int_{V_{cell}}\int_{V_{cell}} n_1 \frac{\rho_2}{4\pi r_{l_2}^2}\xi(r_{12})dV_1 dV_2. \quad (A4)$$

$\rho_g$ is the background emissivity originating from the catalogued galaxies themselves, and so $\eta_p$ is the intensity contributed by these galaxies to the background (which may arise from just a fraction of them). $\eta_c$ arises from the clustering of the galaxies with the background sources. The integrals are over $V_{cell} = \omega \int_z dv(z)$. $r_l$ and $r_{l_2}$ are luminosity distances and $r_{12}$ is the physical separation between $dV_1$ and $dV_2$. Distances and volumes are defined in Appendix B for different cosmologies.

## A1 The Poisson term

The background we are concerned with in this paper is the X-ray background. If the galaxies have been observed at frequency $\nu$ and $L_x(L_\nu)$ is the potential X-ray luminosity of a galaxy with luminosity $L_\nu$, the volume emissivity of the galaxies in X-ray may be written:

$$\rho_g(z) = (1+z)^{-\beta_X+1}\int_{L_{min}(z)}^{L_{max}(z)} L_x(L_\nu)\phi(L_\nu,z)dL_\nu, \quad (A5)$$

where $\phi(L_\nu,z)$ refers to the galaxy luminosity function at $\nu$ and $z$, and $\beta_X$ is their spectral index in the X-ray band. $L_{min}(z)$ and $L_{max}(z)$ depend on the flux limits of the survey and the spectral index of the sources at $\nu$. The mean X-ray intensity contributed by the catalogued galaxies is:

$$\Delta \bar{I}_g = \int_z \frac{\rho_g(z)}{4\pi r_l^2(z)}dv(z) \quad (A6)$$

and therefore:

$$\eta_p = \omega \Delta \bar{I}_g. \quad (A7)$$

## A2 The clustering term

If we assume that sources at significantly different redshifts are uncorrelated ($\xi \equiv 0$ when $|r_2 - r_1| >> 0$), we may do the following change of variables, in *comoving* coordinates:

$$\begin{cases} X_c = \frac{r_{c_1}+r_{c_2}}{2} \approx r_{c_1} \approx r_{c_2} \\ X_l = (1+z)X_c \\ u = r_{c_2} - r_{c_1} \\ dV_1 dV_2 = F^2 X_c^4 dX_c d\Omega_1 d\Omega_2 du \end{cases} \quad (A8)$$

where $F = \frac{dV/d\omega}{r_c^2 dr_c}$ (=1 in flat space), as defined in Appendix B. The clustering term becomes:

$$\eta_c = \int\int n(z) \frac{\rho_s(z)}{4\pi X_l^2} \xi(r_{12}, z) dV_1 dV_2 \quad (A9)$$

$$= \int n(z) \frac{\rho_s(z)}{4\pi X_l^2} X_c^4 F^2 dX_c \int\int d\Omega_1 d\Omega_2 \int \xi(r_{12}, z) du$$

In the small angle approximation, the *physical* separation between the 2 sources is:

$$r_{12}^2 \approx (u^2 F^2 + X_c^2 \theta^2)/(1+z)^2 \quad (A10)$$

### A2.1 power spectrum formalism

$$\xi(r, z) = \frac{1}{2\pi^2} \int_0^\infty P(k,z) \frac{\sin(kr/a)}{(kr/a)} k^2 dk \quad (A11)$$

where $k$ is the *comoving* wavenumber and $a = (1+z)^{-1}$ is the scale factor. The integral over $u$ is:

$$\int_{-\infty}^{+\infty} \xi(r_{12}, z) du = 2 \int_0^\infty \xi(r_{12}, z) du$$

$$= \frac{1}{\pi^2} \int_0^\infty \int_0^\infty P(k,z) \frac{\sin(kr/a)}{(kr/a)} k^2 \, dk \, du$$

$$= \frac{1}{\pi^2} \int_0^\infty P(k,z) \, k \, dk \int_0^\infty \frac{\sin(kr/a)}{(kr/a)} k \, du \quad (A12)$$

The inside bit is:

$$\int_0^\infty \frac{\sin(kr/a)}{(kr/a)} k \, du = \int_0^\infty \frac{\sin(k^2u^2F^2 + k^2X_c^2\theta^2)^{1/2}}{(k^2u^2F^2 + k^2X_c^2\theta^2)^{1/2}} k \, du$$

$$= \int_0^\infty \frac{\sin(t^2 + k^2X_c^2\theta^2)^{1/2}}{(t^2 + k^2X_c^2\theta^2)^{1/2}} F^{-1} dt$$

$$= F^{-1} \frac{\pi}{2} J_0(kX_c\theta) \quad (A13)$$

where $J_0$ is a Bessel function defined as:

$$J_0(x) = \frac{2}{\pi} \int_0^\infty \frac{\sin(t^2+x^2)^{1/2}}{(t^2+x^2)^{1/2}} dt = \frac{2}{\pi} \int_0^\infty \sin x(\cosh u) du.$$

Inserting this into $\eta_c$, we can write:

$$\eta_c = \int_0^\infty P(k,z) k \tilde{g}_0(k) dk \quad (A14)$$

where the kernel $\tilde{g}_0$ is defined as:

$$\tilde{g}_0(k) = \frac{1}{2\pi} \int_c n(z) \frac{\rho_s(z)}{4\pi X_l^2(z)} X_c^4(z) F(z) dX_c(z)$$

$$\times \int_\omega \int_\omega d\Omega_1 d\Omega_2 J_0(kX_c\theta) \quad (A15)$$

Similar calculations have been presented by Baugh & Efstathiou (1994) for the galaxy auto-correlation.

### A2.2 power-law approximation

$$\xi(r,z) = (1+z)^{-(3+\epsilon)} \left(\frac{r}{r_o}\right)^{-\gamma} \quad (A16)$$

The integral over $u$ becomes:

$$\int_{-\infty}^{+\infty} \xi(r_{12}, z) \, du$$

$$= (1+z)^{-(3+\epsilon)+\gamma} r_0^\gamma \int_{-\infty}^{+\infty} (u^2 F^2 + X_c^2 \theta^2)^{-\gamma/2} du$$

$$= (1+z)^{-(3+\epsilon)+\gamma} r_0^\gamma F^{-1} H_\gamma (X_c \theta)^{1-\gamma} \quad (A17)$$

where $H_\gamma$ is the usual stuff in $\Gamma$ functions:

$$H_\gamma = \int_{-\infty}^{+\infty} dx (1+x^2)^{-\gamma/2} = \frac{\Gamma(\frac{1}{2})\Gamma(\frac{\gamma-1}{2})}{\Gamma(\frac{\gamma}{2})} \quad (A18)$$

For a square beam with size $\omega = a^2$, the integral over $\theta$ can be written (Totsuji & Kihara 1969):

$$\int_\omega \int_\omega d\Omega_1 d\Omega_2 \theta^{1-\gamma} = C_\gamma a^{5-\gamma} \quad (A19)$$

Finally,

$$\eta_c = a^{5-\gamma} H_\gamma C_\gamma r_0^\gamma \times \quad (A20)$$

$$\int_z n(z) \frac{\rho_s(z)}{4\pi X_l^2(z)} (1+z)^{-(3+\epsilon)+\gamma} X_c^{5-\gamma}(z) F(z) dX_c(z)$$

For $\gamma = 1.8$, $H_\gamma = 3.68$ and $C_\gamma = 2.25$.

## A3 Angular correlation functions

The background/background and galaxy/galaxy *auto-correlation functions* and the galaxy/background *cross-correlation function* at angular separation $\theta$ are respectively defined as:

$$\begin{array}{l} \eta_{bb}(\theta) = <\delta I_{\theta'} \delta I_{\theta'+\theta}> = <I_{\theta'} I_{\theta'+\theta}> - \bar{I}^2 \\ \eta_{gg}(\theta) = <\delta N_{\theta'} \delta N_{\theta'+\theta}> = <N_{\theta'} N_{\theta'+\theta}> - \bar{N}^2 \\ \eta_{gb}(\theta) = <\delta N_{\theta'} \delta I_{\theta'+\theta}> = <N_{\theta'} I_{\theta'+\theta}> - \bar{N}\bar{I} \end{array} \quad (A21)$$

There is no Poisson terms this time since the cells do not overlap ($\theta > 0$), simply:

$$\begin{array}{l} \eta_{bb}(\theta) = \int_z \frac{\rho_{x_1}}{4\pi r_{l_1}^2} \frac{\rho_{x_2}}{4\pi r_{l_2}^2} dV_1 \int \xi(r_{12}, z) dV_2 \\ \eta_{gg}(\theta) = \int_z n_1 n_2 dV_1 \int \xi(r_{12}, z) dV_2 \\ \eta_{gb}(\theta) = \int_z n(z) \frac{\rho_s(z)}{4\pi X_l^2(z)} dV_1 \int \xi(r_{12}, z) dV_2 \end{array} \quad (A22)$$

With the same change of variables as in the previous section (Eq. A8), we can write:

$$\begin{array}{l} \eta_{bb}(\theta) = \int_0^\infty P(k,z) k \tilde{g}_{bb}(k\theta) dk \\ \eta_{gg}(\theta) = \int_0^\infty P(k,z) k \tilde{g}_{gg}(k\theta) dk \\ \eta_{gb}(\theta) = \int_0^\infty P(k,z) k \tilde{g}_{gb}(k\theta) dk \end{array} \quad (A23)$$

where

$$\begin{array}{l} \tilde{g}_{bb}(k\theta) = \frac{1}{2\pi} \int_z \left(\frac{\rho_s(z)}{4\pi X_l^2}\right)^2 J_0(kX_c\theta) X_c^4 F dX_c \\ \tilde{g}_{gg}(k\theta) = \frac{1}{2\pi} \int_z n^2(z) J_0(kX_c\theta) X_c^4 F dX_c \\ \tilde{g}_{gb}(k\theta) = \frac{1}{2\pi} \int_z n(z) \frac{\rho_s(z)}{4\pi X_l^2} J_0(kX_c\theta) X_c^4 F dX_c \end{array} \quad (A24)$$

or else, in the power-law approximation:

$$\eta_{bb}(\theta) = H_\gamma r_0^\gamma \theta^{1-\gamma} \int_z \left(\frac{\rho_s(z)}{4\pi X_l^2}\right)^2 (1+z)^p X_c^{5-\gamma} F \mathrm{d}X_c$$
$$\eta_{gg}(\theta) = H_\gamma r_0^\gamma \theta^{1-\gamma} \int_z n^2(z)(1+z)^p X_c^{5-\gamma} F \mathrm{d}X_c \quad (A25)$$
$$\eta_{gb}(\theta) = H_\gamma r_0^\gamma \theta^{1-\gamma} \int_z n(z)\frac{\rho_s(z)}{4\pi X_l^2}(1+z)^p X_c^{5-\gamma} F \mathrm{d}X_c$$

where $p = \gamma - (3 + \epsilon)$ and again $X_c(z)$ is the comoving distance coordinate $r_c(z)$ and $X_l(z)$ is the corresponding luminosity distance defined in Appendix B. Here we have assumed the size of the cells to be infinitesimal $\mathrm{d}\omega$, so that the previous integrals $\int_\omega \int_\omega \mathrm{d}\Omega_1 \mathrm{d}\Omega_2 J_0(kX_c\theta)$ or, in the power-law approximation, $\int_\omega \int_\omega \mathrm{d}\Omega_1 \mathrm{d}\Omega_2 \theta^{1-\gamma}$, over the cell size are replaced in the above equations by $\mathrm{d}\omega_1 \mathrm{d}\omega_2 J_0(kX_c\theta)$ and $\mathrm{d}\omega_1 \mathrm{d}\omega_2 \theta^{1-\gamma}$ respectively. All the above correlation functions are thus implicitly defined per unit square solid angle.

## APPENDIX B: COSMOLOGY

In the following, $r_c$ is the comoving radial coordinate in units of Mpc and $\mathrm{d}v = \mathrm{d}V/\mathrm{d}\omega$ the volume element per unit solid angle in units of $\mathrm{Mpc}^3 \mathrm{sr}^{-1}$. $F$ is a curvature function defined as $F = \mathrm{d}v/r_c^2 \mathrm{d}r_c$ (=1 in flat space). The luminosity distance in Mpc is defined as $r_l(z) = (1+z)r_c(z)$.

### B1  $\Omega_0 = 1$ and $\Lambda = 0$ : Flat (Einstein–de Sitter)

$$r_c(z) = (2c/H_0)\left[1 - (1+z)^{-1/2}\right] \quad (A26)$$

$$F(z)\frac{\mathrm{d}r_c}{\mathrm{d}z}(z) = \frac{\mathrm{d}v}{r_c^2 \mathrm{d}z} = \frac{(c/H_0)}{(1+z)^{3/2}} \quad (A27)$$

### B2  $\Omega_0 < 1$ and $\Lambda = 0$ : Open (Friedmann–Robertson–Walker)

$$r_c(z) = (2c/H_0)\frac{\Omega_0 z + (\Omega_0 - 2)[(1+\Omega_0 z)^{1/2} - 1]}{\Omega_0^2(1+z)} \quad (A28)$$

$$F(z)\frac{\mathrm{d}r_c}{\mathrm{d}z}(z) = \frac{\mathrm{d}v}{r_c^2 \mathrm{d}z} = \frac{(c/H_0)}{(1+z)(1+\Omega_0 z)^{1/2}} \quad (A29)$$

### B3  $\Omega_0 < 1$ and $\Lambda = 3(1-\Omega_0)H_0^2$ : Flat with a Cosmological Constant (Friedmann–Lemaître)

$$r_c(z) = (c/H_0)\int_0^z \frac{\Omega_0^{-1/2} \mathrm{d}z'}{[(1+z')^3 - 1 + \Omega_0^{-1}]^{1/2}} \quad (A30)$$

$$F(z)\frac{\mathrm{d}r_c}{\mathrm{d}z}(z) = \frac{\mathrm{d}v}{r_c^2 \mathrm{d}z} = \frac{(c/H_0)}{\Omega_0^{1/2}[(1+z)^3 - 1 + \Omega_0^{-1}]^{1/2}} \quad (A31)$$